\documentclass[a4paper,twocolumn]{esapub2005}
\usepackage{epsfig}
\newcommand{\var}{\rm{Var}}

\begin{document}

\title{Extraction of the photometric information : corrections}
  \author[1] {R. Samadi} 
\author[1]{F. Fialho}
\author[1]{J.E.S. Costa} 
\author[1]{V. Lapeyrere} 
 \author[2]{R. Drummond}  
 \author[1]{L. Pinheiro  Da Silva} 
 \author[3]{F. Baudin} 
  \author[3]{P. Boumier}

\affil[1]{Laboratoire d'Etudes Spatiales et d'Instrumentation pour l'Astrophysique, Observatoire de
Paris, 92190, Meudon, France. }
\affil[2]{Instituut  voor Sterrenkunde, KULeuven, Leuven, Belgium}
\affil[3]{Institut d'Astrophysique Spatiale (IAS), b\^atiment 121, F-91405 Orsay (France), 
Universit\'e Paris-Sud 11 and CNRS (UMR 8617) }
\affil[4]{Laboratoire d'Astrophysique de Marseille, BP 8, 13376 Marseille cedex 12 
(France), Universit\'e de Provence,  CNRS (UMR 6110) and CNES}

\maketitle

\begin{abstract}
We present here the set of corrections that will be applied to the raw
data of the CoRoT mission. The aim is to correct
the data for instrumental and environmental perturbations
as well as to optimize the duty-cycle, in order to reach the
expected performances of the mission. 
The main corrections are : the correction of the electromagnetic
interferences, suppression of outliers, the background correction, the jitter 
correction and the correction of the integration time variations.
We focus here on these corrections and emphasize their efficiency.
\end{abstract}

\section{Introduction}

There will be different levels of corrections performed on the
scientific data. 

The first level will be
applied on the raw data and deals with first order correction 
associated with  instrumental and environmental (e.g. background) perturbations.
These will be based on calibrations performed on the
ground before and during the launch as well as on-board.
The goal of such first level corrections is to reach the expected
performances in terms of noise and orbital components.

The first order corrections will be applied throught  the
so-called   N0-N1 pipeline \footnote{``N'' refers to ``Niveau'' in
  french, level in english. The N0 data corresponds to the raw data.}.

Corrections of higher levels will be performed in order to optimize
the global performances of the mission, in particular in order 
to optimize the duty-cycle of the mission and to remove residual
instrumental and environmental perturbations 

The second order corrections will be applied throughout the so-called N1-N2 pipeline. 

The first pipeline will be operated every one to seven days and
therefore deals with short term corrections. The
second pipeline will be operated once a run is finished.

These proceedings are organized as follows: 
In Sect.~\ref{N0N1} we describe the content of N0-N1 pipeline and in
Sect.~\ref{N1N2} that of the N1-N2. The different corrections applied in the N0-N1 pipeline are detailed in Sections ~\ref{offset} to \ref{jitter}.

\section{N0-N1 pipeline}
\label{N0N1}

The corrections applied to the star light-curves (LC
hereafter) in the N0-N1 pipeline are the following:
\begin{itemize}
\item offset subtraction (see Sect.~\ref{offset})
\item suppression of the outliers (see Sect.~\ref{outliers})
\item correction of the electromagnetic interferences (EMI, see Sect.~\ref{emi})
\item gain correction (to transform digitized data into electrons,
  see Sect.~\ref{gain})
\item integration time correction (see Sect.~\ref{integration time variations})
\item background subtraction (see Sect.~\ref{background})
\item jitter correction (see Sect.~\ref{jitter})
\end{itemize}

Before performing these treatments we need to carry out the offset
 and background measurements.
For the offset measurements, the treatments are the following:
\begin{itemize}
\item correction of electromagnetic interference (see Sect.~\ref{emi}) 
\item suppression of the outliers (see Sect.~\ref{outliers})\\ 

 For the background measurements, we must apply in addition
 the following treatments:
\item offset subtraction 
\item gain correction (see Sect.~\ref{gain})
\item calibration of the sky background aacross the field of view in order to
   remove {\it a posteriori} the background contribution to the star LCs (see Sect.~\ref{background}).
\end{itemize}

Each of the above-mentioned treatments is detailed in Sects~\ref{offset}-\ref{jitter}.
\subsection{Offset}
\label{offset}

The offset of the electronic chain is measured on-board for each time
step (1s for the astero channel and 32s for the exo-planet
channel). 

For the astero channel the offset is subtracted from the
star and background LCs  on-board. However the offset is subject to
electromagnetic interferences (see Sect.~\ref{emi}) that will only be
corrected afterwards on ground. Moreover, although these offset measurements are
corrected on-board for outliers (e.g. cosmic impacts or glitches),
the residual outliers will be removed on ground (see Sect.~\ref{outliers}).  
We must then go back to the offset subtraction performed on-board and
subtract instead the offset treated on ground.

For the exoplanet channel, the electronic offset is not subtracted
 from the star and background LCs on-board. 
The offset measurements are processed on ground to correct them
for EMI and cosmic impacts (or glitches). Afterwards, they are
subtracted from the LCs of the exoplanet channel.

\subsection{Outliers}
\label{outliers}

For the exo channel, the on-board measurements (background and
stars) are not corrected for the outliers (e.g. cosmic impacts or
glitches). This correction is applied on ground.

For the astero channel, although the measurements are
corrected for outliers on-board, this correction may not be fully efficient - in
particular when the satellite crosses the South Atlantic Anomaly. 
This is why the measurements from the astero channel are re-processed on
ground in order to correct them for residual cosmic impacts (or
glitches).

Several standard algorithms will be applied in order to correct the LC
for outliers.

\subsection{Electromagnetic interferences}
\label{emi}

\subsubsection{Description and consequences}

The charge transfer and readout performed on one CCD induces
electromagnetic interferences (EMI hereafter)  on all the three other
CCDs (see Pinheiro da Silva {\it et al}, this volume).

\begin{figure*}[ht]
   \begin{center}
     \epsfig{file=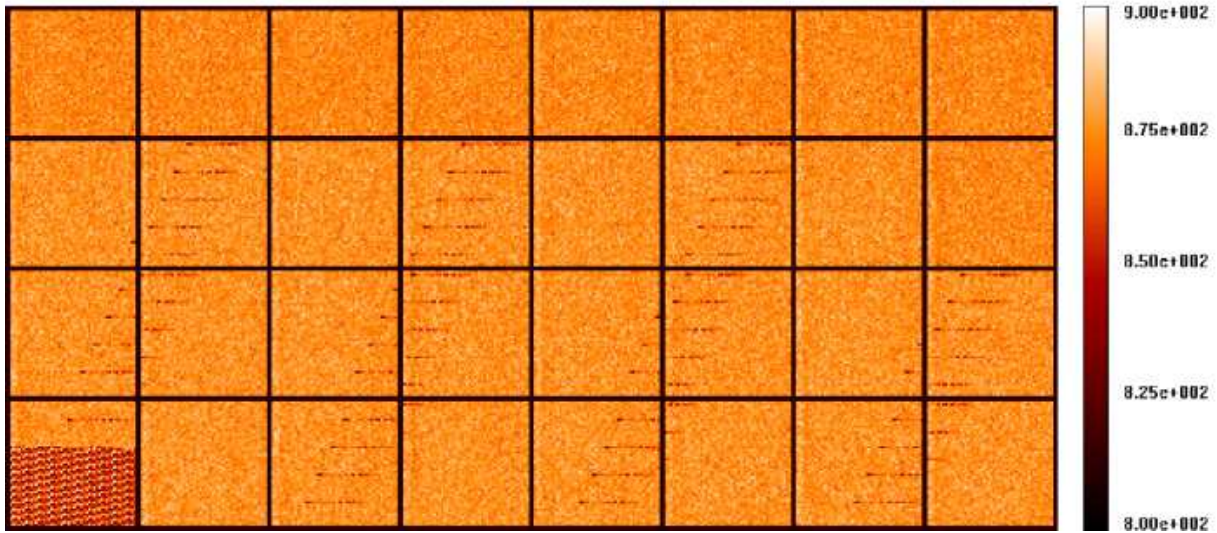, width=16cm}
  \end{center}
\caption{The pattern induced by the EMIs on the seismo CCD. 
Each box corresponds to an exposure of 1s; 32 exposures
  are represented.}
\label{emi:fig3}
\end{figure*}

On the exoplanet channel, during a 32s integration cycle, the
first 23s are used to transfer and readout the charges. During
the 9 remaining seconds, no more readouts are performed and the
electronic chain does not interfere with the other CCDs anymore.
On the astero channel, the charge transfer and readout are performed
periodically each second and take less than 1s.
As a consequence, the EMIs induced by the CCDs of the
exoplanet chains on the CCDs of the astero channels occur during the
first 23 exposures of a 32s cycle and cease during the 9 remaining exposures.
The amplitudes and shapes of these EMIs depend on the position of the
 astero target being perturbed by the exoplanet CCDs.
However these perturbations are strictly periodic (period of 32s).  
The  patterns of the EMIs seen on the CCD of
the seismo channel are shown in Fig.\,\ref{emi:fig3}

\begin{figure}[ht]
   \begin{center}
     \epsfig{file=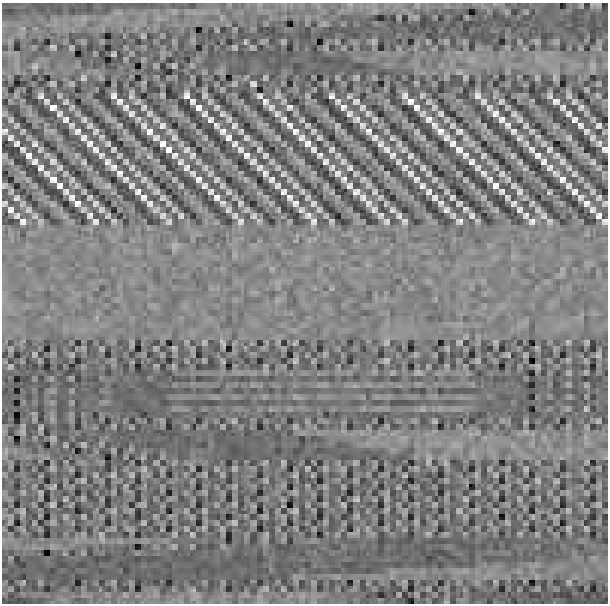, width=8cm}
  \end{center}
\caption{The pattern induced by the EMIs on the exo CCD. }
\label{emi:fig4}
\end{figure}

In the same way the charge transfer and readout performed on the seismo
CCD induce EMIs on the exo  CCD. The  patterns corresponding to
those EMIs  are shown in Fig.\,\ref{emi:fig4}. Their shapes depend on
the position of the astero targets.   

\subsubsection{Effects on the photometry}

\begin{figure}[ht]
   \begin{center}
     \epsfig{file=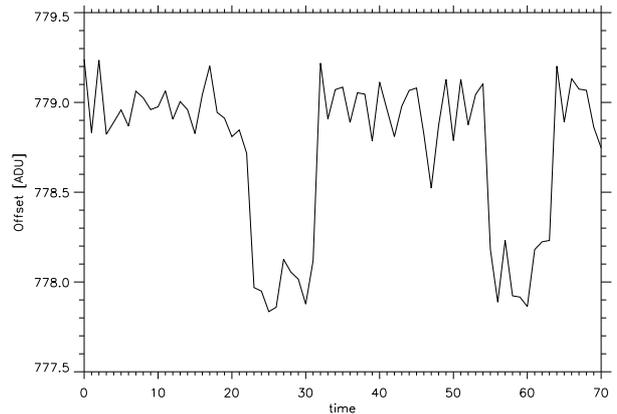, width=8cm}
   \end{center}
\caption{Offset of the electronics as measured on the astero CCD.}
\label{emi:fig1}
\end{figure}

The perturbations induced by the EMIs on the offset measured on the
astero channel are illustrated in Fig.~\ref{emi:fig1}.
As seen in the Fig. the first 23 exposures of a 32s cycle have a
larger offset level than the 9 remaining exposures.
On the other hand, the EMI induced by the CCDs of the
astero channels  on the CCDs of the exoplanet channels occur each
second and the associated perturbations are constant in
time for the exposures of the exo channel.

 Their amplitudes and shapes also depend on the
positions of the astero targets on the astero CCDs. 

For both channels the perturbations induced by the EMIs on the LCs
(star, background) or the offset measurements are equivalent
to a shift of the electronic offset with respect to the nominal
 level of the electronic offset.
For the astero channel this shift varies with time and depends on the
position of the target on the CCD.
For the exoplanet channel however, this shift is constant with time
but is inhomogeneous in space and hence differs from a target to another.

Such shifts of the electronic offset introduce biases on the absolute
values of the measured target fluxes. However the absolute value of the
target flux is not relevant for the main scientific goals of the
mission.
Nevertheless, if we are interested in the absolute
values of the measured fluxes, such corrections are required.
In addition, for the astero channel, as the perturbations induced by the
EMIs change during a 32s cycle, their variations with time are
undesirable as it decreases the efficiency of the outlier corrections.

%, we can accommodate with such biases.
%However for the astero channel, as the perturbations induce by the EMIs change during a 32
%cycle, their variations with time are undesirable unless the exposures are binned by
%32.
%The latter  binning is out of order (? JUSTIFIER)! for the first grade data (the so called N1
%data, produced  after the first level of corrections).

\subsubsection{Corrections}

The shapes and positions of the EMIs on the CCD can be predicted precisely from
the known electronic micro sequences that govern the charge transfers
and readouts. However their amplitudes can only be characterized
through calibrations.

{\it \noindent Exo channel:}

To correct the perturbations induced by the EMIs on
exoplanet LCs we first calibrate the EMIs perturbations 
expected for each different electronic micro-sequences (see Pinheiro
da Silva {\it et al}, this volume). 
>From this calibration and according to the way the
windows of the astero CCDs are settled,  we derive next 
the EMIs patterns as shown in Fig.\,\ref{emi:fig4}.
Then, for each target, we integrate the piece of the pattern that falls in
the target template. This gives us a value of the shift
induced by the EMIs on the LC for each target.
Again, as the amplitude of the EMIs motif can change on a large scale, we
 can  repeat the calibration and the correction processes periodically.

{\it \noindent Astero channel:}

For the Astero channel we can either perform  a correction similar to
the one applied for the exo-channel or   perform an empirical correction.

In the first method, we begin by calibrating the EMIs perturbations 
expected for each different electronic micro-sequence.
From this calibration we derive, for each window target, the
expected perturbations for all phases of a 32s cycle (see Fig.~\ref{emi:fig1}).
 These perturbations depend on the way the
windows of the astero CCDs are programmed.
We next integrate the piece of the pattern that falls in
the target mask.  This gives us a value of the shift
induced by the EMIs on the LC for each target

\begin{figure}[ht]
   \begin{center}
     \epsfig{file=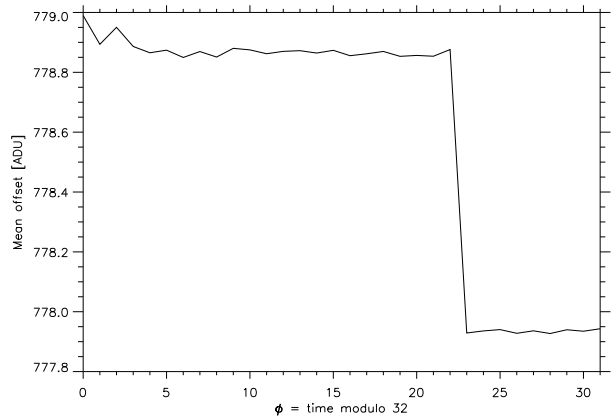, width=8cm}
   \end{center}
\caption{Offset of the electronic (Fig.~\ref{emi:fig1}) averaged at
  constant phase $\phi$ (see text).}
\label{emi:fig2}
\end{figure}

In the second method, we  proceed as follows:
As the perturbations are periodic with a period of 32s, we average the
offset sequence at constant phase $\phi$, where the phase $\phi$ of the offset
sequence is by definition $\phi \equiv$~t~modulo~32 where $t$ is the time
(in seconds, arbitrary origin). The result is shown in
Fig.~\ref{emi:fig2} for offset measurements. As seen in the Fig., the offset stands at a high
level during 23 seconds and then remains at a lower level during the
remaining 9 seconds of the 32s cycles. Once the motif of the EMI
perturbation is calibrated in this way, we apply this motif on the
offset sequence to correct it.
However, as the amplitude of the motif can change on
larger time scales, we split the offset sequence in an adequate
number of sub-series. We then perform the calibration and the
correction on each sub series.

\subsection{Gain}
\label{gain}

The gain of the electronic chain and of the CCD have been measured on
ground during the calibration of the camera and the electronics.
All the lightcurves of both channels will be multiplied by the gain in order to
convert ADU into electrons. 

During the mission, periodic calibrations of the gain (electronics
+CCD) will be performed. 
However, the quality of those calibrations will not be high enough to derive the
absolute value of the gain. 
However, they will allow us to characterize the variation of the
gain during the life of the instrument and to apply afterwards a gain correction
that will take into account its long term variation.

\subsection{Integration time variations}
\label{integration time variations}

The CCD charge transfer and readout is controlled by a
clock inside the instrument (Quartz Instrument, QI hereafter). 
 Thermal control is performed on the QI which prevents it from
undesirable large fluctuations. 
Therefore variations in the QI with respect to the Universal Time
(UT) are expected to be small (less than $\sim$ 1 ppm per second).

However, each 32s the charge transfer is synchronised with respect
to the clock located inside the Proteus platform (Quartz Platform, QP hereafter).
As a consequence the first 31 exposures are controlled by the QI while
the (last) 32nd exposure is controlled by the QP.  

This QP is not thermally controlled
and hence is subject to significant variations with the  
temperature of the platform. The latter is expected to vary by a
few kelvin during an orbit and the temperature coefficient of the QP is
expected to be of the order of $\sim$ 2 ppm/K. Therefore variations of
QP frequency of the order of few ppm are expected during an orbit.

Consequently, the integration time of the first 31 exposures is
expected to remain almost constant while that of the 32nd exposure is
expected to vary by $\sim 10^{-4}$ second if we assume a variation
of the platform temperature of $\Delta T \simeq$ 2 K.
This clock variation induces a flux variation at the orbital frequency
of the order of $\sim 4$ ppm for an observation of 5 days. This is two times
larger than the requirements. 

Fortunately, each 32s the platform will provide, through the
telemetry, the GPS (Global Positioning System) pulses received by the
platform and the values of the QP counters when those pulses are received.
>From this information, we will then be able to measure the variation of
the QP with respect to the UT. 
In turn, this will allow us to correct the variation of the
integration time of the 32nd exposures.
This correction consists of multiplying the flux of the 32nd
exposures by the ratio of the nominal integration time \footnote{the
  1s nominal sampling time minus the
duration of the charge transfer (0.206s), gives a  nominal integration time
of $\sim$ 0.794s} and the actual duration of the  32nd
exposures (the latter being derived from the measured variations of
the QP). 

This correction  efficiently removes the perturbations induced by the
variations of the QP on the astero LCs.
 Furthermore it is so small that it doesn't significantly change the
 (Poisson) statistics of the charge collected by the CCD. 

For the exoplanet channel, the  CCD charge transfer and readout are
synchronised with respect to the QP. Therefore the flux measured on
the exoplanet channel will also suffer from the
variations of QP. The level of the flux perturbation induced by the
QP variations will be of the same level as that expected for the
astero channel. Thus it will be below the requirements imposed on
the exoplanet channel. Nevertheless, this flux perturbation will be corrected
in the same manner.

\subsection{Background}
\label{background}

The correction for sky background is a standard procedure in 
the reduction process of photometric data and basically consists of
the subtraction of the sky background level from the photometric
measurements of the star. 

{\it CoRoT Background Windows} --- 
To measure the sky background light, CoRoT has {\it background windows} 
distributed over each CCD, as shown in the Tab.\ref{tab:bg_windows}. 
The background windows on the CCDs of the SEISMO channel are larger, 
but binned. The modes {\it oversampled} (OV) and {\it non-oversampled} (NOV)
of the EXO channel have background windows of the same size, 
but different in number per CCD.

% TABLE --- BACKGROUND WINDOWS OVER EACH CCD.
\begin{table}
  {\scriptsize
  \begin{tabular}{lccccc}
    Channel   & CCD &  Number of               & Size          & Binning     & Sampling \\
              &     &      windows             & (in pix)      & (in pix)    & (in sec) \\  \hline
    SEISMO     & A1  &  5                       & $50\times 50$ & $5\times 5$ & 1   \\
              & A2  &  5                       & $50\times 50$ & $5\times 5$ & 1   \\ \hline
    EXO (OV)  & E1  &  10                      & $10\times 10$ & --          & 32  \\
              & E2  &  10                      & $10\times 10$ & --          & 32  \\ \hline
    EXO (NOV) & E1  &  40                      & $10\times 10$ & --          & 512 \\
              & E2  &  40                      & $10\times 10$ & --          & 512 \\ \hline
  \end{tabular}}
  \caption{CoRoT background windows.}
\label{tab:bg_windows}
\end{table}
% ---------

\subsubsection{Background Components}

The background is made up of different components. The major
contributors are the diffuse galactic background,
the zodiacal light, the straylight from the Earth and the South
Atlantic Anomaly (SAA). Other minor contributors are the moonlight,
the parasite light due to diffusion and reflection of the 
stars' light inside the instrument.

While the diffuse galactic background and the zodiacal light are
almost constant over time, the Earth straylight and the SAA are
periodic components and require special attention.

\begin{figure}[h]
   \begin{center}
     \epsfig{file=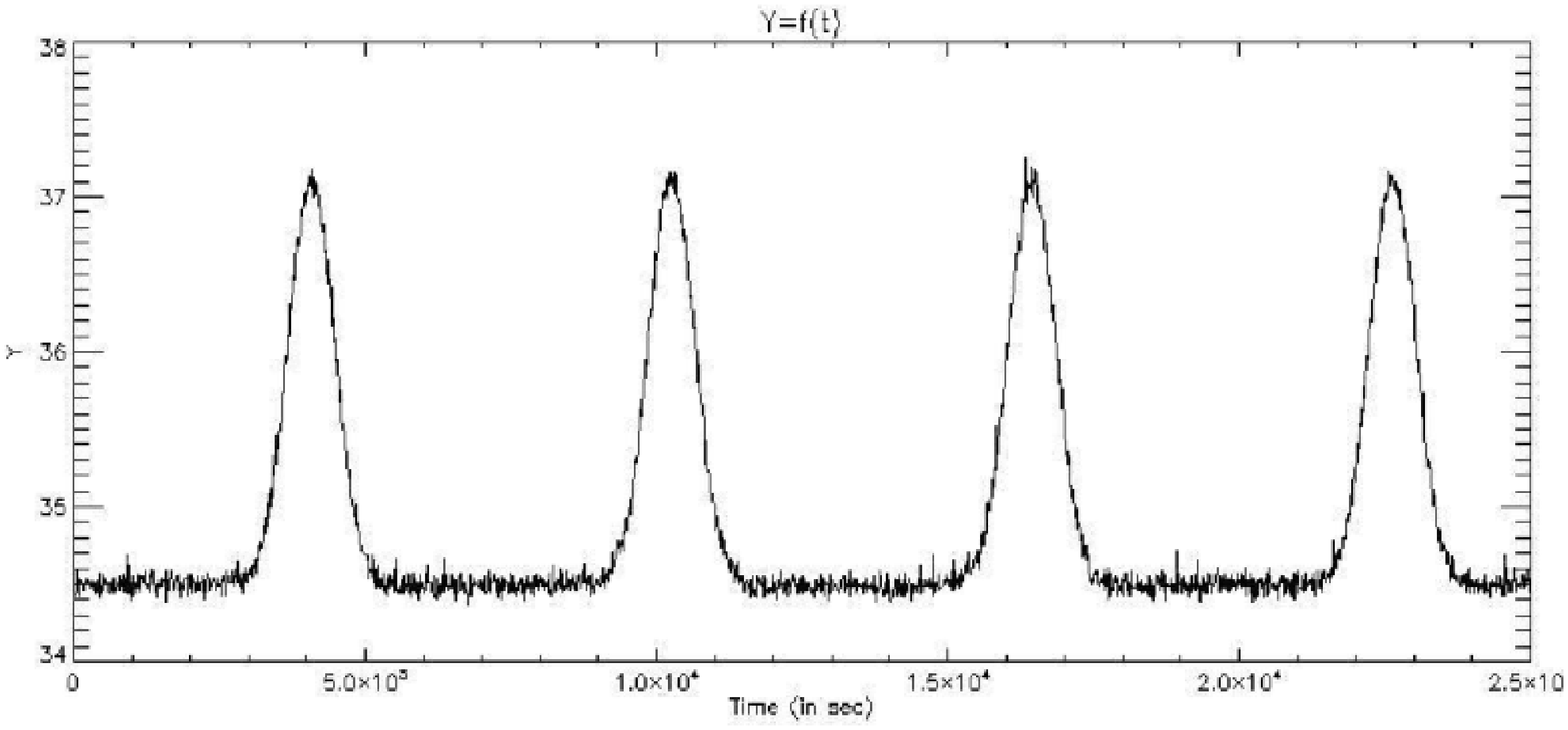, width=8cm, angle=0.}
   \end{center}
\caption{Earth straylight light curve. The bumps occur when the
satellite passes over the illuminated face of the Earth.}
\label{fig:bg_straylight}
\end{figure}

\begin{figure}[ht]
   \begin{center}
     \epsfig{file=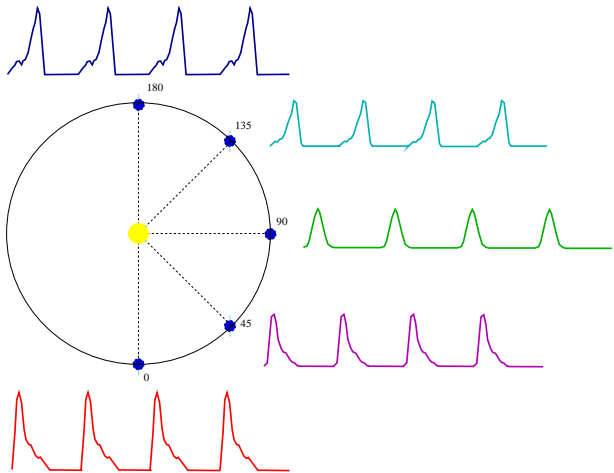, width=8cm}
   \end{center}
\caption{The curve of the Earth straylight changes as a function of the
         orbital position of the Earth around the Sun.}
\label{fig:bg_orbit}
\end{figure}

{\it Earth Straylight } --- 
The level of incident Earth straylight increases when the satellite
passes over the illuminated face of the Earth and is at a minimum when it
passes over the dark face, as illustrated in Fig.\ref{fig:bg_straylight}. 
The height and shape of the bumps depend on the orientation of
the satellite orbital plane in relation to the Sun-Earth axis and the
worst case occurs when the two are aligned as shown in
Fig.\ref{fig:bg_orbit}.

% FIGURE -- EARTH STRAYLIGHT LIGHT CURVE
% ----------

% FIGURE -- EARTH STRAYLIGHT vs. ORBITAL MOTION
% ----------

\subsubsection{Effects on the photometry}

\begin{figure}[ht]
   \begin{center}
     \epsfig{file=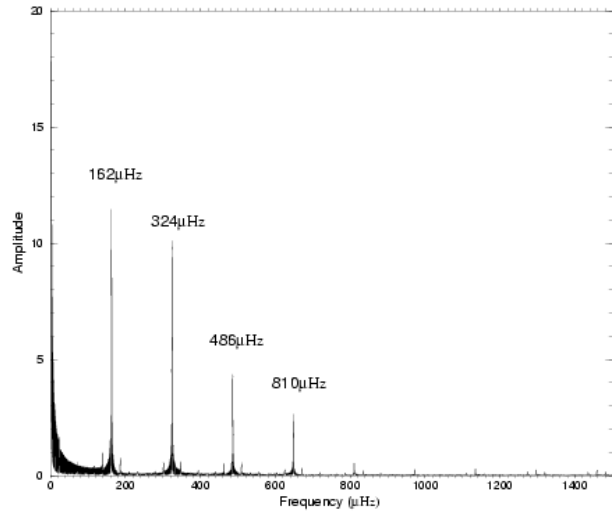, width=8cm}
   \end{center}
\caption{Discrete Fourier transform of the Earth straylight curve.
         The fundamental frequency is $162\,\mu Hz$.}
\label{fig:bg_dft}
\end{figure}

The Earth straylight introduces a periodic signal with period equal to orbital period (1h 43min) in the measured light curve of the stars. 
As shown in Fig.\ref{fig:bg_dft} in
the Fourier spectrum, it appears as a peak with the fundamental frequency
of $162\mu Hz$ and a series of harmonic frequencies, with low amplitude,
but that can trouble the detection and analysis of pulsation frequencies
of the observed pulsating stars.

% FIGURE -- DFT OF THE EARTH STRAYLIGHT
% ---------

Earth straylight can also affect the detection
of planetary transits in the EXO channel. If a planetary transit occurs
while the satellite is passing over the illuminated face of the Earth,
the increase in the background light level can compensate for the small
variations in the light curve of the star.

\subsubsection{Sky Background Gradient}
\begin{figure}[ht]
   \begin{center}
     \epsfig{file=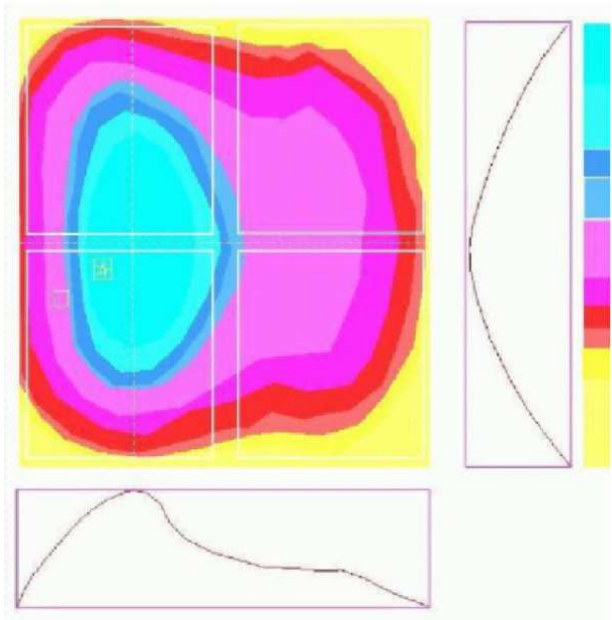, width=8cm}
   \end{center}
\caption{Theoretical diffusion of the sky background light over
         the focal plane.}
\label{fig:bg_gradient}
\end{figure}

Due to the asymmetry of the CoRoT optical system, the diffusion of
the sky background light over the focal plane is non-homogeneous.
We don't know exactly if the non-homogeneity is regular or irregular
and how strong the gradient is. However, theoretical studies suggest
a regular gradient as shown in Fig.\ref{fig:bg_gradient}.

% FIGURE -- GRADIENT
% ---------

\subsubsection{Correction}

The simplest procedure for correction by sky background of a star
light curve is to subtract from each measurement of the star light
the sky level measured from the {\it nearest} background window. 
This procedure assumes that the background is homogeneous, at least,
in the neighborhood of the two windows.
However, in presence of a non-negligible gradient, the sky levels
for the two windows can be significantly different and this
procedure will not completely subtract the sky from the star
light curve.

In order to avoid this problem, we need take into account the
sky background gradient.
For this, the gradient is modeled by
a spatial function $f(x,\,y)$ over the whole CoRoT focal plane,
from the combined sky measurements of all CCDs.
Different trial functions are fitted and the one that provides the
best fit is chosen as model for the background gradient.

In the fitting, it is necessary
to take into account that the {\it gains} of the channels 
are different.
For this reason we introduce the {\it gain} parameters, $g_{ij}$,
where the index $i$ refers to the CCD (A1, A2, E1, E2) and 
the index $j$ refers to the CCD's channel (left or right).
In fact, the correction by gain is done in a previous step of
the data treatment, using values
experimentally determined. This way, the determination of the
parameters $g_{ij}$ is used to refine the gain correction.

\begin{figure}[ht]
   \begin{center}
     \epsfig{file=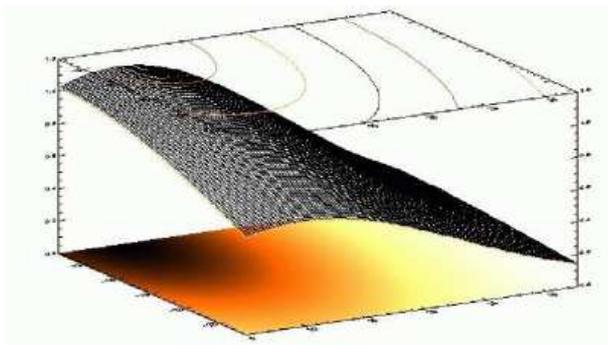, width=8cm}
   \end{center}
\caption{Fitting of a Gaussian model for $f(x,\,y)$ from simulated data.}
\label{fig:bg_fitting}
\end{figure}

The trial functions used are listed in Tab.\ref{tab:bg_trailFunctions}. 
Functions of higher order can be tried, if needed.
Fig.\ref{fig:bg_fitting} shows the result of a fitting of a
Gaussian model for the background gradient from simulated data.

% FIGURE -- GAUSSIAN GRADIENT
% ---------

\begin{figure}[ht]
   \begin{center}
     \epsfig{file=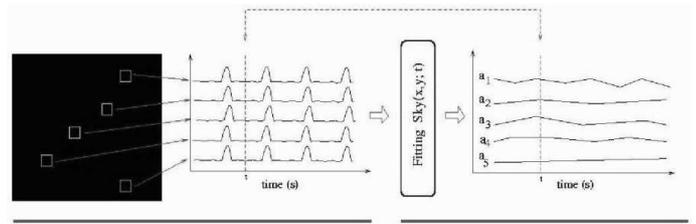, width=9cm}
   \end{center}
\caption{From the measurements of the sky background obtained in the instant
         $t$ a spatial function $Sky(x,\,y)$ is fitted and its parameters are
         calculated. By doing this for each time step, a parameters time series is obtained.}
\label{fig:bg_param}
\end{figure}

As a result, a time series is obtained for each one of the parameters
of the model function which allow us to calculate, by interpolation with respect
to time, the correct sky level over each star window and for any moment in time
(see Fig.\ref{fig:bg_param}).

% TABLE - TRAIL FUNCTIONS
\begin{table}
\begin{center}
{\scriptsize
  \begin{tabular}{l|l|c}
    Gradient      &          & Number of  \\
    Type          & Function & Parameters            \\ \hline
    Constant      & $f(x,\,y) = g_{ij}\,a_0$     & 9 \\
    Linear        & $f(x,\,y) = g_{ij}\,(a_0 + a_1\,x + a_2\,y)$ & 11 \\
    Quadratic     & $f(x,\,y) = g_{ij}\,(a_0 + a_1\,x + a_2\,y$  & \\
                  & $+ a_3\,xy + a_4\,x^2 + a_5\,y^2)$             & 14 \\
    Gaussian      & $f(x,\,y) = g_{ij}\,\left[a_0\,\exp\left(  a_1 (x-a_2)^2 \right.\right.$   & 13 \\
                  & $\left.\left.+ a_3 (y-a_4)^2\right)\right]$ &  \\  \hline
  \end{tabular}
}
 \caption{Trial functions for modeling of the sky background gradient.}
\end{center}
  \label{tab:bg_trailFunctions}
\end{table}
% --------

% FIGURE -- EARTH STRAYLIGHT
% ---------

The modeling of the sky background as described above assumes that the
gradient is non-homogeneous, but relatively regular. This hypothesis
can be tested {\it a posteriori} from the comparison of the variance for
the fitting $\var_{\rm{mod}}$ with the average variance of the background
measurements, $\var_{\rm{bg}}$:

% TABLE -- STATISTICAL TEST
\begin{center}
{\scriptsize
\begin{tabular}{ccl}
comparison  &  gradient & procedure to be used \\ \hline
$\var_{\rm{mod}} < \var_{\rm{bg}}$       & non-homogeneous,  & model $f(x,\,y;\,t)$ \\
                                            & but regular       &                       \\ \hline
$\var_{\rm{mod}} \simeq \var_{\rm{bg}}$  & homogeneous      & $f(x,\,y;\,t)=\rm{constant}$  \\
                                            &                   & \underline{or} the nearest sky window \\ \hline
$\var_{\rm{mod}} > \var_{\rm{bg}}$       & non-homogeneous   & local triangularization      \\
                                            & and irregular     &                        \\ \hline

\end{tabular}}
\end{center}
% -----------

\begin{figure}[h]
   \begin{center}
     \epsfig{file=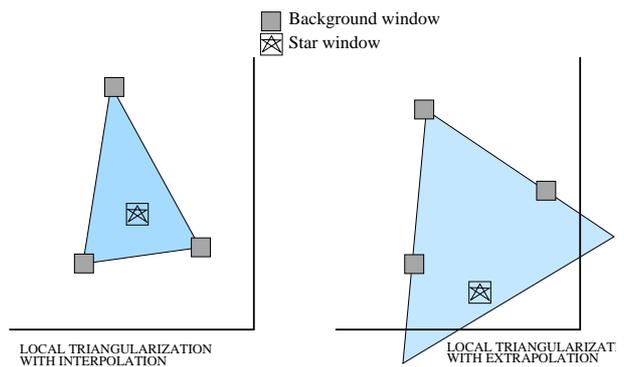, width=8cm}
   \end{center}
\caption{The technique of local triangularization used to calculate the
         background level over a sky window from the measurements of the
         three nearest sky windows.}
\label{fig:bg_triang}
\end{figure}

For non-homogeneous and irregular gradients,
the calculation of the sky level over the star window by {\it local triangularization} is used.
This procedure calculates the parametric equation of the plane defined by the  measurements 
of the {\it three} nearest background windows and then calculates
the background level for the star window position.
This technique is illustrated in Fig.\ref{fig:bg_triang}.

% FIGURE -- TRIANGULARIZATION
% ----------

\subsection{Jitter}
\label{jitter}

The idea here is to correct the LCs for the edge effects and/or chromatic
contamination provoked by satellite jitter in the fixed photometric aperture
in both channels.
The importance of this correction is not only seen in the improvement of
the S/N ratio of the photometric curves, but also in better definition of
seismo photometric masks, because the quality we can achieve in jitter
correction is also taken into consideration when calculating the optimal mask.
See item VI.3.
As the seismo and exo channels have different goals, their designs are not
the same, which means that different approaches can be used to
conceive the corrections. Below the proposed methods to correct each
channel are described.

\subsubsection{Seismo channel}

For the seismo channel the CoRoT instrument was specified in a way
that the jitter photometric noise is negligible for stars of magnitude 6. \
However the correction of this source of noise must be carried out for less bright stars, or if the AOCS degrades. 

\begin{figure}[ht]
   \begin{center}
     \epsfig{file=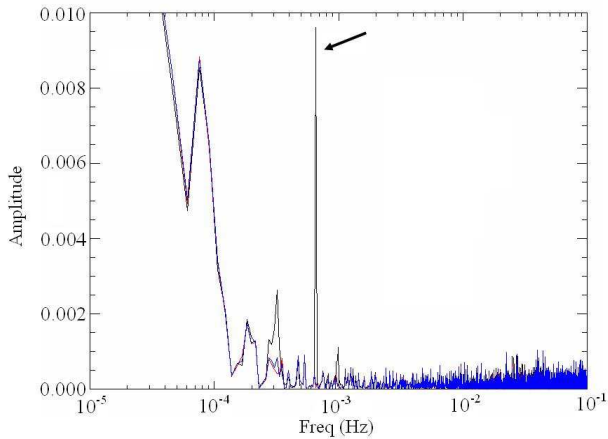, width=8cm}
   \end{center}
\caption{Fourier spectrum of non corrected (black) and corrected (blue) LCs}
\label{jit:fig1}
\end{figure}

Fig.\,\ref{jit:fig1} illustrates the problem in frequency. The component of frequency
indicated by the arrow is due to jitter and must be corrected.

\begin{figure}[ht]
   \begin{center}
     \epsfig{file=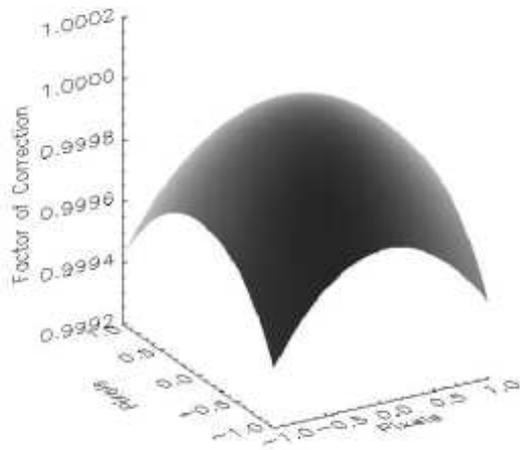, width=8cm}
   \end{center}
\caption{Example of correction surface for jitter correction in seismo
field }
\label{jit:fig2}
\end{figure}

Two methods of correction have been developed for this channel with a single
idea behind them: to estimate the best correction according to the
equation: 
\begin{equation} F_{c} = K(\Delta x_i,\Delta y_i) F_{m},
\end{equation}
where $Fc$ is the corrected flux,  $K (\Delta x_i,\Delta y_i)$ is the
so called correction surface, $Fm$ is the measured flux and
$\Delta x_i,\Delta y_i$ are the displacement of the target with respect
to its average position. 
This means that once created, the correction surface in Fig.\,\ref{jit:fig2}, 
can be used to apply the adequate factor of correction to each measured point of the
photometry as a function of the pointing error at that instant.

Below we describe the two ways this surface can be created.

{\it \noindent PSF Model:}
	
This is a model-based method that consists of creating a surface based on
the evaluation of loss of flux from the Point Spread
Function (PSF) model and the photometric aperture. The PSF is gotten at the beginning of life of the satellite (see Pinheiro da Silva {\it et al}, this volume).

{\it \noindent Signal correlation:}

By examining the correlation among calculated displacement (x and y jitter) 
and output (measured flux), we estimate the average loss of
flux for a pre-defined jitter variation grid. We average over a period of more than 10 orbits
to build the correction surface from the COROT signal itself. This needs no accurate knowledge of the PSF.
For a more detailed description of the various methodologies see \cite{dru06}.
\subsubsection{Exo channel}

\begin{figure}[ht]
   \begin{center}
     \epsfig{file=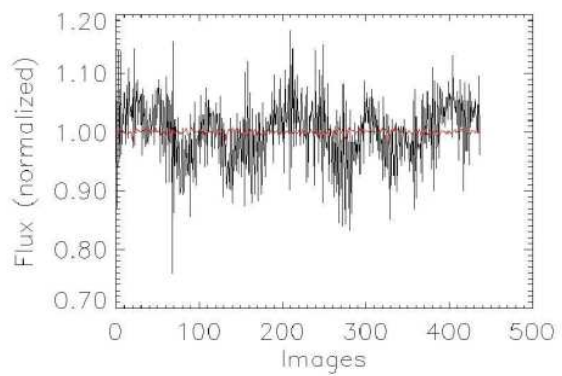, width=8cm}
   \end{center}
\caption{Exo LCs before (black) and after (red) jitter correction
}
\label{jit:fig3}
\end{figure}

The seismo channel is almost insensitive to the variation of
jitter, but the exo channel is highly sensitive to the pointing
variation of the satellite. This is true not only because of the lower
number of pixels in the mask, but also because of the use of chromatic
photometry that induces chromatic contaminations. As an example, the Fig.\,\ref{jit:fig3}
shows two time series curves before and after correction, where we can see a
significant gain in the SNR.

	Again, as in the seismo channel, we are interested in estimating the
correction factor $K (\Delta x_i,\Delta y_i)$. Two methods are considered here and they are explained
below.

{\it \noindent PSF Model:}

\begin{figure}[ht]
   \begin{center}
     \epsfig{file=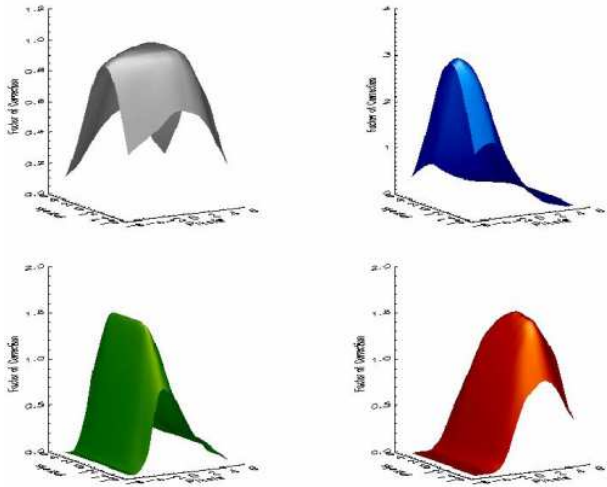, width=8cm}
   \end{center}
\caption{Correction surfaces for mono and chromatic jitter correction in
exo field}
\label{jit:fig4}
\end{figure}

	This is a simple extension of the method applied to the seismo channel.
Instead of creating just one correction surface, four surfaces are
calculated, one for each color. The difference in each is the mask it
uses. Fig.\,\ref{jit:fig4} illustrates the surfaces for each color.

{\it \noindent Integrated spectrum:}

\begin{figure}[ht]
   \begin{center}
     \epsfig{file=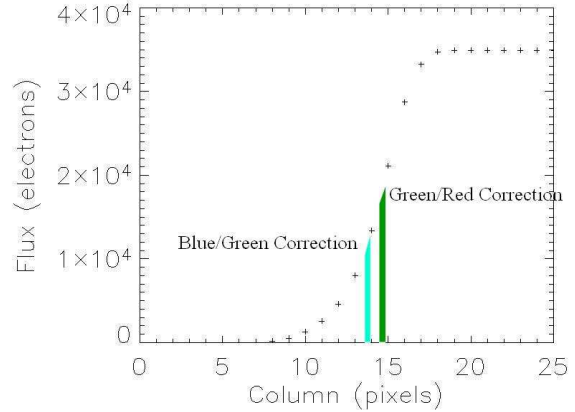, width=8cm}
   \end{center}
\caption{Star integrated spectrum for chromatic jitter
correction in exo field}
\label{jit:fig5}
\end{figure}

	This method is based on the integrated spectrum of the star (Fig.\,\ref{jit:fig5}
that is simply the summation of all pixels in one direction (columns in this
case). Here it is not necessary to have an accurate knowledge of the PSF. A few
images are enough to compute the integrated spectrum. As the chromatic
effects are the most significant in exo photometric apertures, only this
phenomenon is evaluated. This means that here we are just interested in 1-D
correction and ithis is done by the means of the estimation of the flux loss 
in the chromatic borders of the integrated spectrum. Chromatic borders are
calculated using energy thresholds. Thus, it is possible to recover the
photometric signal lost as a function of jitter. This method demands a larger mask
to guarantee that edge effects will be negligible.

\section{N1-N2 pipeline}
\label{N1N2}

The corrections performed by the N1-N2 pipeline have not yet been
completely specified, since we will need the real data to do this.

They are different for the exo and astero channels.
We describe below the current content of this pipeline, successively for the astero
 channel (Sect.~\ref{N1N2_astero}) and the exo channel (Sect.~\ref{N1N2_exo}).

\subsection{Astero channel}
\label{N1N2_astero}

We will first process the so-called ``imagettes''. These are 
small images of a star that will be downloaded every 32s for some of the
 targets. A  Point Spread Function (PSF) fitting will be applied to
 those ``imagettes'' in order to: 
\begin{itemize}
\item  extract  the photometry of the star
\item detect cosmic impacts 
\item derive the sky background
\end{itemize}
For bright stars ($m_v \sim 6$), the photometry based on PSF
fitting is shown to be less accurate than the aperture photometry
performed on-board. The former is only useful for fainter
stars ($m_v \sim 8$) or  when the satellite crosses the SAA.

On the star LCs, we will first transform the Universal Time (UT) into the
barycentric time (BT hereafter) in order to move to a  time reference that has a
constant Doppler shift with respect to the time reference relative to the star.
 
Furthermore, in the on-board reference time (UT), the sampling is almost constant while in
the  barycentric time reference, this is no longer the case. 
This is a problem if one performs a Fast Fourier Transform (FFT) because the
FFT assumes that the measurements are  evenly sampled.
In order to perform a FFT, we interpolate the LC with respect to an evenly sampled BT grid.

Finally, we integrate the LC over 32s. 

When crossing the SAA, the photometry extracted from the 
``imagettes'' on the basis of a PSF fitting will be more accurate. Hence,
during the SAA crossing, we will patch the LC with the photometry of
the  ``imagettes''. This procedure will increase the duty-cycle. 

If the sky background derived from the ``imagettes'' turns out to
be less biased than the one subtracted by the N0-N1 pipeline (see
Sect.~\ref{emi}), we will cancel the sky background correction
performed by the the N0-N1 pipeline and subtract instead the one
derived from the ``imagettes''.  

Finally, the correction of the long term variation of the gain will be performed on the complete
LC (see Sect.~\ref{gain}).

\subsection{Exo channel}
\label{N1N2_exo}

The exo channel N1-N2 pipeline will be divided in two parts.

The first part will be devoted to additional instrumental corrections
which are not fully taken into account in the N0-N1 pipeline.
Two corrections are foreseen at the moment: (i) a correction for
residual cosmic events near the SAA, and (ii) a correction for
residual background signal.
The correction for cosmic impacts envisaged in the N0-N1 pipeline
(see Sect.~\ref{outliers}) is indeed expected to leave residuals when the
satellite is at the edge of the SAA, before exo treatments stop. The aim of the background correction in the N1-N2 pipeline is to complement the N0-N1 background
correction of Sect.~\ref{background} in case the background signal is inhomogeneous
across the field-of-view.
The algorithms to be applied still need to be defined.
Additional residual corrections may be implemented during the mission
if new instrumental effects are identified.

The second part will be devoted to a rough characterization of the
level of stellar variability, in the form of a numerical parameter
which will be included in the header of the data. The aim of this
parameter will be to help to select the best candidates where
low level transit events can be searched for.
A method based on a Fourier analysis of the lightcurve will be
implemented at the beginning of the mission. This method will be refined later
on and/or complemented by new indicators of stellar
variability.

%\begin{acknowledgements}
{\it  The content of this paper represents the work of all the CoRoT ground
segment team in Meudon (LESIA) and in Orsay (IAS). }
%\end{acknowledgements}

\end{document}